\documentclass[prl,aps,twocolumn,groupedaddress,showpacs,floatfix]{revtex4}
\usepackage{amsmath,amssymb,multirow,epsfig,bm}

\newcommand{\etal}{{\it et al.,\;}}
\newcommand{\beq}{\begin{equation}}
\newcommand{\eeq}{\end{equation}}
\newcommand{\bea}{\begin{eqnarray}}
\newcommand{\eea}{\end{eqnarray}}

\newcommand{\veps}{\varepsilon}

\newcommand{\benn}{\begin{displaymath}}
\newcommand{\eenn}{\end{displaymath}}

\begin{document}

\title{Thermodynamics of a Trapped Unitary Fermi Gas }

\author{ Aurel Bulgac$^1$, Joaqu\'{\i}n E. Drut$^1$ and Piotr Magierski$^{2}$ }
\affiliation{$^1$Department of Physics, University of
Washington, Seattle, WA 98195--1560, USA}
\affiliation{$^2$Faculty of Physics, Warsaw University of Technology,
ulica Koszykowa 75, 00-662 Warsaw, POLAND }

\begin{abstract}
 
We present the first model-independent comparison of recent measurements of the entropy and of the critical temperature of a unitary Fermi gas, performed by Luo~{\it et al.}, with the most complete results currently available from finite temperature Monte Carlo calculations. The measurement of the critical temperature in a cold fermionic atomic cloud is consistent with a value $T_c=0.23(2)\veps_F$ in the bulk, as predicted by the present authors in their Monte Carlo calculations. 

\end{abstract}

\date{\today}


\maketitle


The study of the properties of a Fermi gas in the unitary regime (when the 
$s$-wave scattering length $a$ is large compared to the average interparticle separation) emerged as one of the most fascinating theoretical many-body problems since it was first formulated by G.F. Bertsch as the Many-Body X (MBX) challenge in 1999 \cite{gfb,baker}. 
The experimental investigation of the unitary Fermi gas (UFG) began with its realization in cold atomic traps by O'Hara {\it et al.}~at Duke University three years later \cite{jet}. At unitarity (often referred to as ``at resonance"), when $a\rightarrow\infty$, the properties of such a system are governed by deceptively simple laws. In particular, the ground state energy per particle is given by $E/N=3 \veps_F\xi/5$, where $\veps_F = \hbar^2k_F^2/2m$ is the Fermi energy of a noninteracting Fermi gas with the same number density $n=N/V=k_F^3/3\pi^2$. The determination of the dimensionless constant $\xi$ was the subject of the MBX challenge and the best current accepted value was determined a bit later through restricted/fixed node Monte Carlo (MC) calculations as $\xi = 0.42(1)$ \cite{carlson,chang,reddy,giorgini}. This value was confirmed by the zero temperature extrapolation of unrestricted MC calculations of Ref. \cite{bdm}, where $\xi=0.44(3)$ was obtained. Theoretically, it was also found that this system is superfluid at low temperatures and the value of the pairing gap was estimated at zero temperature to be $\Delta = 0.504(24)\veps_F$ \cite{carlson,chang,reddy}. (For lack of space we quote and comment here only on theoretical results obtained in controlled MC calculations, where the errors are typically, though not always, only of statistical origin. In all other theoretical approaches that we are aware of, the errors are essentially impossible to quantify due to the lack of any identifiable small parameter.) A number of finite temperature thermodynamic properties of the homogeneous phase was determined as well 
\cite{bdm,burovski,dl,trivedi}, even though there is still some disagreement concerning the exact value of the critical temperature $T_c$, on which we shall comment later. The temperature dependence of the pairing gap has not been determined yet. On the experimental side there is a quite wide spread in values of the dimensionless parameter $\xi$ \cite{note} determined in various experiments. However, the latest experiments seem to converge, possibly guided by the existence of firm theoretical results, to the expected value: 0.74(7) \cite{jet}, 0.51(4) \cite{kinast}, $0.32^{+0.13}_{-0.10}$ \cite{xi_grimm}, 0.36(15) \cite{xi_ens2004}, 0.46(5) \cite{xi_rice}, 0.45(5) \cite{xi_jila}, 0.41(15) \cite{xi_ens2007}. 
The measurements of the pairing gap are still in their infancy.
Although it has been conclusively demonstrated that a UFG is superfluid  at sufficiently low temperatures \cite{vortices}, the value of the pairing gap has only been determined so far in one experiment \cite{gap_exp}. Moreover,
the extracted value is significantly smaller than the theoretical value \cite{carlson,chang,reddy}. 
Such a small pairing gap is inconsistent with the value of $T_c$ measured independently \cite{kinast,luo}. In Ref. \cite{luo}, $T_c$ was found by first determining the entropy and the energy of such a system, a procedure which allows to establish an absolute temperature scale. 
In this experiment a number of properties of the atomic cloud were determined in the unitary regime. The cloud was then adiabatically brought to the BCS side of the Feshbach resonance, where $k_Fa=-0.75$ (here $k_F$ is the Fermi momentum corresponding to the central density of the cloud). In this regime $(k_F|a|<1$) one can use many-body perturbation theory \cite{fw}, together with the Local Density Approximation (LDA), to evaluate various cloud properties (see Ref. \cite{giorgini} for a comparison of MC results with perturbative many-body results). Measurements of the energy of the cloud can thus be related, using theory, to temperature and entropy. Since the entropy is conserved, by measuring the energy of the cloud one can determine the energy-entropy dependence in the unitary regime and its absolute temperature as well (from $T=\partial E(S,N)/\partial S$). 
Most of the atomic trapping potentials used in these experiments can be approximated rather well with harmonic potential wells. Such potentials can be shown to satisfy the virial theorem at unitarity, namely $E(T,N)=2 N\langle U\rangle = 3m\omega_z^2\langle z^2\rangle$ \cite{virial}, and therefore simply measuring the spatial shape of the cloud allows for a unique determination of the UFG energy at any temperature. One of the main goals of the present work is to present a critical analysis of the results of this experiment \cite{luo}, in the light of available finite temperature MC calculations. This work thus represents the first model-independent comparison of a full theory directly with experiment.

\begin{figure}
\epsfxsize=9cm
\centerline{\epsffile{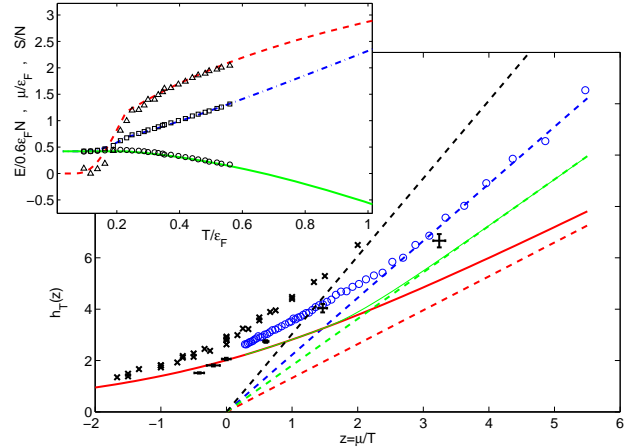}}
\caption{ \label{fig:pressure}  MC data from Ref.\cite{bdm} (blue circles), Ref.\cite{burovski} (six black points) and Ref. \cite{dl} (black crosses). The four straight lines starting at the origin are the $T\rightarrow 0$ limits of $h_T(z\rightarrow\infty)=2^{2/5}z/\xi^{3/5}$, where $\xi=0.25(2)$ \cite{dl}, 
$\xi=0.42$ \cite{reddy,giorgini}, $\xi= 0.59$ for meanfield/BCS approximation and 
$\xi=1$ for the free Fermi gas model respectively. The two solid lines (red/lower and green/higher) correspond to $h_T(z)$ calculated in the free Fermi gas and the BCS/meanfield approximation $h_T(z)$ respectively.  In the inset we show the fits to MC data \cite{bdm}. }
\end{figure}

  
At unitarity $(1/k_Fa=0)$ the pressure of a homogeneous UFG is determined by a universal function $h_T(z)$: 
\beq
\mathcal{P}(T, \mu ) 
        = \frac{2}{5}\beta\left [ T h_T\left ( \frac{\mu}{T}\right )\right ]^{5/2}\! , \quad \!\!\! 
\beta = \frac{1}{6\pi^2}\left(\frac{2m}{\hbar^2}\right)^{3/2},
\eeq
where $T$ and $\mu$ are the temperature and the chemical potential, respectively. Remembering that the grand canonical potential is $\Omega(V,T,\mu)=-V\mathcal{P}(T, \mu )$ one can easily show that 
the energy of the system reads: $E=3\mathcal{P}V/2$, where $V$ is the volume
of the system. Following a reasoning \cite{forbes} similar to that in Ref. \cite{BulgacForbes} one can show that thermodynamic stability implies positivity $h_T(z)\geq 0$ and convexity $h_T''(z)\geq 0$. In the high-temperature limit $\mu \rightarrow -\infty$ and $\mathcal{P}(T, \mu )$ tends from above to the free Fermi gas pressure. In the low-temperature limit $\mathcal{P}(T, \mu )$ tends from above to $\mathcal{P}(0, \xi \veps_{F} )=4\beta \veps_F^{5/2}\xi/5$. 
Similarly, at all temperatures the pressure calculated in the BCS/meanfield approximation will give a variational estimate from below of $\mathcal{P}(T,\mu)$. In Fig. \ref{fig:pressure} we illustrate these statements and plot the results of three finite temperature MC calculations \cite{bdm,burovski,dl}. It is not possible to extract the data for this plot from Ref. \cite{trivedi}. The results of Ref. \cite{dl} stand apart from the rest of the theoretical and experimental results. While the results \cite{bdm} agree with these bounds, three out of the six calculated points in Ref. \cite{burovski} (one of the lowest $T$/highest $z$ and two at the highest $T$/lowest $z$'s) slightly violate them.  The point at $\mu/T\approx 3.24$ is approximately where the authors of Ref. \cite{burovski} claim that the normal-superfluid phase transition occurs.

We use our MC results \cite{bdm} (with a spatial lattice size $8\times 8 \times 8$) to generate smooth interpolation formulas for the energy, chemical potential and entropy (see inset of Fig. \ref{fig:pressure}). Standard manipulations show that all the UFG thermodynamic potentials can be expressed in terms of a single function of one variable, a property known as universality \cite{bdm,burovski,ho}. This property was incorporated in our interpolation. At high temperatures we notice that our results smoothly approach the corresponding free Fermi gas results with some offsets for the energy, chemical potential and entropy \cite{bdm}. 

At this point we assume that the LDA can be used to describe the properties of an atomic cloud in a trap. There has been no systematic study of the accuracy of LDA in the unitary regime. We can, however, easily estimate the role of the gradient corrections for a noninteracting Fermi gas in an anisotropic harmonic trap. Using methods described in \cite{brack} one can show that the ground state energy of a two-component fermion system in an anisotropic harmonic trap is given by:
\beq
E(N) = \frac{ \hbar\Omega(3N)^{4/3}}{4}+
       \frac{ \hbar\overline{\omega}(3N)^{2/3}}{8} [1 + e_{sc}(N)],
\eeq
where $\Omega=(\omega_x\omega_y\omega_z)^{1/3}$ and 
$\overline{\omega}=(\omega_x^2+\omega_y^2+\omega_z^2)/3\Omega$. In this formula the first term is the leading LDA contribution. Naively, one would expect the next to leading order LDA correction to be proportional to $N$. It is a peculiarity of harmonic potentials, however, that the leading gradient corrections start at next order instead, namely at $O(N^{2/3})$ \cite{brack}.  At the same order one finds the so-called shell correction to the energy, given in this formula by $ \hbar\overline{\omega}(3N)^{2/3}e_{sc}(N)/8$. The function $e_{sc}(N)$ has a vanishing average over particle number, is minimum when a shell is filled and maximum in the middle of a shell. This term depends strongly on the asymmetry of the harmonic potential and has maximum amplitude for spherical potentials. In such case the amplitude of $e_{sc}(N)$ is about 0.5, while it is significantly smaller for asymmetric wells. Finite temperatures \cite{brack} and pairing, even at unitarity \cite{george}, have a smoothing effect on the shell energy correction term, but do not affect in any other major way the remaining leading gradient correction term.  At temperatures close to $T_c$ one would expect the gradient corrections to play a noticeable role in the description of the pairing properties \cite{urban}. However, close to $T_c$ the pairing energy represents a relatively small contribution to the total energy, and so the errors in the total energy from using naive LDA around $T_c$ are likely to be small too. In a trap, the fraction of particles that are close to loosing superfluidity (namely for which $x({\bf r}=T/\veps_F({\bf r})) \leq x_c=0.23(2)$, where $x_c$ is the critical temperature in natural units \cite{bdm}) is also small. All in all, it appears that for the mostly-harmonic traps used in typical experiments the role of the gradient corrections is relatively small and LDA is a reasonable approximation. 

In this approach, the grand canonical thermodynamic potential for a UFG confined by an external potential $U(\bf{r})$ is a functional of the local density $n(\bf r)$ given by
\begin{equation}
\Omega = \int dV \left [
\frac{3}{5}\veps_F({\bf r})\varphi (x)n({\bf r}) +
U({\bf r}) n({\bf r})-\lambda n({\bf r}) \right ],
\end{equation}
where 
\begin{equation}
\label{eq:XandEF}
x({\bf r}) = \frac{T}{\veps_F({\bf r})}, \quad 
\veps_F({\bf r}) = \frac{\hbar^2}{2m} [3 \pi^2  n({\bf r})]^{2/3},
\end{equation}
and we have used the universal form for the free energy per particle 
$F/N$ in the unitary regime:
\begin{equation}
\frac{F}{N}=\frac{E-TS}{N} = \frac{3}{5}\veps_F\varphi(x) = 
\frac{3}{5}\veps_F [\xi(x)-x\sigma(x)],
\end{equation}
where for a homogeneous system $\xi (x)= 5E/3\veps_FN$, $\sigma(x)=S/N$ is the entropy per particle and $x = T/\veps_F$  (see inset in Fig. \ref{fig:pressure}). 
The overall chemical potential $\lambda$ and the temperature $T$ are constant throughout the system. The density profile will depend on the shape of the trap as dictated by $\delta \Omega / \delta n({\bf r}) = 0$, which results in: 
\begin{equation}
\label{eq:chempot}
\frac{\delta \Omega}{\delta n({\bf r})}=\frac{\delta(F-\lambda N)}{\delta n({\bf r})}
=\mu(x({\bf r})) + U({\bf r})-\lambda .
\end{equation}
At a given $T$ and $\lambda$, equations (\ref{eq:XandEF}) and (\ref{eq:chempot}) completely determine the density profile $n(\bf r)$ (and consequently both $E(T,N)$ and $S(T,N)$) in a given trap for a given total particle number.  The only experimental input we have used is the particle number, the trapping potential and the scattering length at $B=1200\; G$, taken from Ref. \cite{luo}. The potential was assumed to be an `isotropic' Gaussian, although it is not entirely clear to us to what extent this is accurate, especially in the axial direction. We have approximated the properties of the atomic cloud at $B=840\;G$ with those at unitarity ($B=834\;G$), where we have MC data. For $B=840\;G$ and for the parameters of the Duke experiment \cite{luo} one obtains $1/k_Fa = -0.06$, using data of Ref. \cite{grimm}, if the Fermi momentum corresponds to the central density of the cloud at $T=0$.

\begin{figure}
\includegraphics[width=9cm]{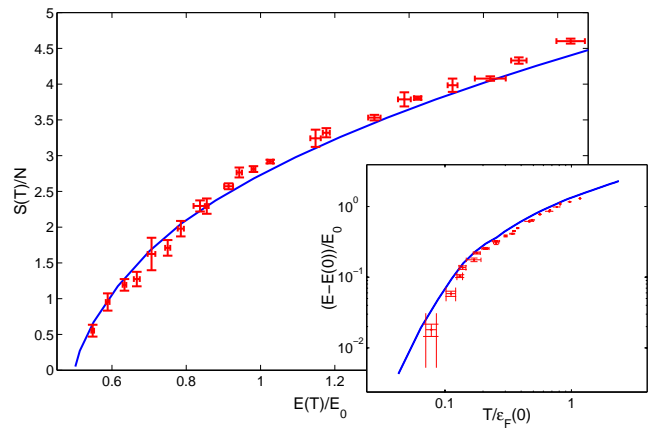}
\caption{\label{fig:SofE} Entropy as a function of energy for the UFG in the Duke trap \cite{luo}: experiment (points with error bars) and present work (solid curve), where $E_0=N\veps_F^{ho}$. Inset: loglog plot of $E(T)$ as results from our calculations and as derived from experimental data \cite{luo}. The temperature is in natural units, namely the Fermi energy at the center of the trap: $\veps_F(0)$.}
\end{figure}
\begin{figure}
\includegraphics[width=9.0cm]{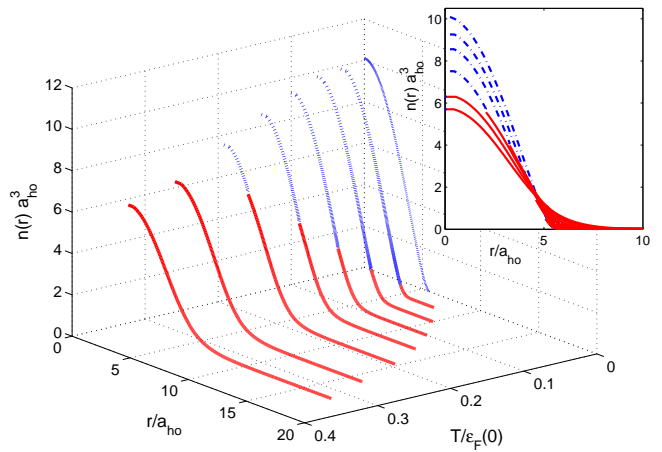}
\caption{\label{fig:profiles} The radial (along shortest axis) density profiles of the Duke cloud at various temperatures, as determined theoretically in the LDA using the MC results of Ref. \cite{bdm}. The dotted blue line shows the superfluid part of the cloud, for which $x({\bf r})=T/\veps_F({\bf r})\leq 0.23$. The solid red line shows the part of the system that is locally normal. Here $a_{ho}^2=\hbar /m\omega_{max}$.}
\end{figure}

Our results for the entropy of the cloud and the density profiles for several temperatures, are shown in Figs. \ref{fig:SofE} and \ref{fig:profiles}. In all the figures the temperature is measured in natural units of $\veps_F(0)$, corresponding to the actual central density of the cloud at that specific temperature. In Refs. \cite{kinast,luo} the temperature is expressed in units of the Fermi energy at $T=0$ in a harmonic trap: $\veps_F^{ho}=\hbar\Omega (3N)^{1/3}$. It is clear from Fig. \ref{fig:profiles} that the central density decreases with $T$ and that the superfluid core disappears at $T_c = 0.23(2)\veps_F(0)$, which translates into $T_c = 0.27(3)\veps_F^{ho}$ to be compared to
$T_c = 0.29(2)\veps_F^{ho}$ of Ref. \cite{luo}. There is a noticeable systematic difference between theory and experiment at high energies, see Fig. \ref{fig:SofE}. This discrepancy can be attributed to the fact that the experiment was performed slightly off resonance, on the BCS side, where $1/k_Fa=-0.06$. Even though theory will soon be extended to this region of $1/k_Fa$, a proper normalization of theory {\it vs.} experiment demands experimental results exactly at resonance.

\begin{figure}[tbh]
\includegraphics[width=8.0cm]{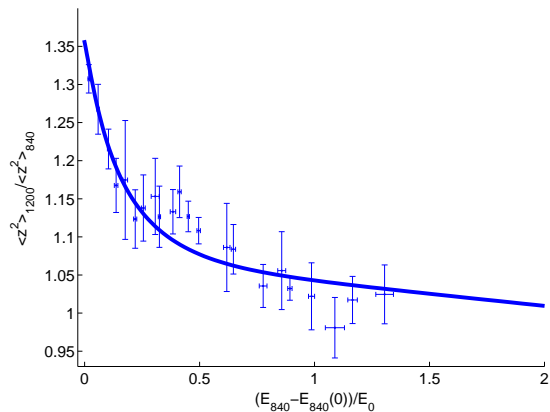}
\caption{\label{fig:zsquared} Ratio of the mean square cloud size $\langle z^2 \rangle$ at $a k_F = -0.75$ ($B=1200\;G$) to its value near unitarity 
($B=840\;G$), as a function of the energy: experiment {\cite{luo}} (points with error bars) and the present theory (solid blue line) .}
\end{figure}

In Fig. {\ref{fig:zsquared}} we show our results for the rms radius, in the form of the ratio of the mean square axial cloud size $\langle z^2 \rangle$ at $k_Fa = -0.75$ to its value at unitarity, as a function of the energy relative to the ground state.  This dependence illustrates the relation between the energies of the cloud (related in turn to the spatial profiles via the virial theorem) at two different values of the magnetic field, but at the same value of the entropy. The quality of the agreement between our theoretical calculations and the experimental data demonstrates the soundness of the entire procedure to determine the entropy and the temperature scale for the UFG. In experiments, one can determine the value $E_c$ of the energy at the transition temperature without knowing the value of the temperature itself, simply by noticing the appearance of a kink in $E$ {\it vs.} $S$ or {\it vs.} ``empirical" $T$ \cite{kinast}. Specifically, it was determined in Refs. \cite{kinast,luo} that $E_c=E(T_c)-E(0)\approx 0.41(5) E_0$, to be compared with what we find theoretically for such a system $E_c=0.32 E_0$. Similarly, it was determined in Ref. \cite{luo} that $S_c=S(T_c)/N\approx 2.7(2)$, to be compared with our result $S_c=2.15$.  
We have identified $T_c$ with the disappearance of the superfluid core, which occurs according to our MC data at $T_c=0.23(2)\veps_F$, 
see Fig. \ref{fig:profiles}. Our MC results are also consistent with a slightly higher $T_c\approx 0.25 \veps_F$, which would lead to $E_c=0.36 E_0$ and $S_c=2.6$ and thus to an almost ``perfect'' agreement between theory and experiment. As mentioned above however, an experiment exactly at unitarity is highly desirable, along with more precise MC data, in order to definitely settle the remaining discrepancies.
The values for the energy $E_c$ and entropy $S_c$ are particularly interesting because their determination does not require knowledge of $T_c$ and can be directly confronted with the MC calculations of Burovski {\it et al.} \cite{burovski}. At low temperatures, the lowest three values of $E(T)$ of 
Ref.~\cite{burovski} agree with our own MC results \cite{bdm}. The three highest temperature points, however, have in our opinion significant systematic errors (see Ref. \cite{bdm} and comments in regards to Fig.~\ref{fig:pressure}). We might safely assume that the energy prediction based on MC data \cite{burovski} should not differ noticeably from ours. However, a
$T_c=0.152(7)\veps_F$ will result in $E_c\approx 0.16 E_0$, in noticeable disagreement with experimental findings \cite{kinast,luo}. Similarly, Ref.~\cite{burovski} determines $S(T_c)/N=0.2(2)$, which leads (according to our calculations for the Duke trap) to $S_c=1.5$, noticeably less than the experimental value. One should notice that an analysis \cite{ts} of the damping of sound modes \cite{sound} is also consistent with $S_c=2.7$.


We are thankful to  B. Clancy, L. Luo and J.E. Thomas, for useful discussions and for providing their experimental data. We also thank Michael McNeil Forbes for extensive discussions, in particular for suggesting the use of the function 
$h_T(z)$. Support is acknowledged from the DOE under grants DE-FG02-97ER41014 and DE-FG02-00ER41132, and from the Polish Ministry of Science. 



\begin{thebibliography}{99}

\bibitem{gfb} G.F. Bertsch, {\em Many-Body X Challenge Problem}, see R.A. Bishop, Int. J. Mod. Phys. {\bf B 15}, {\it iii}, (2001).

\bibitem{baker} G.A. Baker, Jr. Phys. Rev. C {\bf 60}, 054311 (1999).

\bibitem{jet} K.M. O'Hara, {\it et al.}, Science, {\bf 298}, 2179 (2002). 

\bibitem{carlson} J. Carlson \etal Phys. Rev. Lett. {\bf 91}, 050401 (2003).

\bibitem{chang} S.Y. Chang \etal Phys. Rev. A {\bf 70}, 043602 (2004).

\bibitem{reddy} J. Carlson and S. Reddy, Phys. Rev. Lett. {\bf 95}, 060401 (2005).

\bibitem{giorgini} G.E. Astrakharchik \etal Phys. Rev. Lett. {\bf 93}, 200404 (2004).

\bibitem{bdm} A. Bulgac, J.E. Drut, and P. Magierski, Phys. Rev. Lett. {\bf 96} 090404 (2006).

\bibitem{burovski} E. Burovski {\it et al.}, Phys. Rev. Lett. {\bf 96} 160402 (2006); {\it ibidem} {\bf 97}, 239902(E) (2006); New J. Phys. {\bf 8}, 153 (2006).

\bibitem{dl} D. Lee and T. Sh\"affer, Phys. Rev. C {\bf 73}, 015202 (2006); D. Lee, Phys. Rev. B {\bf 73}, 115112 (2006).

\bibitem{trivedi} V.K. Akkineni \etal cond-mat/0608154.

\bibitem{note} Experimentalists quote typically $\beta$, which unfortunately has no physical meaning whatsoever, since all observables are related to $\xi = 1 + \beta$.

\bibitem{kinast} J. Kinast, \etal Science {\bf 307}, 1296 (2005).

\bibitem{xi_grimm} M. Bartenstein \etal Phys. Rev. Lett. {\bf 92}, 120401 (2004).

\bibitem{xi_ens2004} T. Bourdel \etal Phys. Rev. Lett. {\bf 93}, 050401 (2004).

\bibitem{xi_rice}G.B. Partridge \etal Science {\bf 311}, 503 (2006); 

\bibitem{xi_jila} J.T. Stewart \etal Phys. Rev. Lett. {\bf 97}, 220406 (2006).

\bibitem{xi_ens2007} L. Tarruell \etal cond-mat/0701181v1.

\bibitem{vortices} M.W. Zwierlein, {\it et al.},  Nature, {\bf 435 }, 1047 (2005).

\bibitem{gap_exp} C. Chin \etal Science {\bf 305}, 1128 (2004).

\bibitem{luo} L. Luo {\it et al.}, cond-mat/0611566.

\bibitem{fw} A.L. Fetter and J.D. Walecka, {\it Quantum Theory of Many-Particle Systems}, Dover (2003).

\bibitem{virial} J.E. Thomas, J. Kinast, and A. Turlapov,  Phys. Rev. Lett. {\bf 95} 120402 (2005).

\bibitem{forbes} Michael McNeil Forbes, private communication.

\bibitem{BulgacForbes} A. Bulgac and M. M. Forbes Phys. Rev. A {\bf 75}, 031605(R) (2007).

\bibitem{ho} T.-L. Ho, Phys. Rev. Lett. {\bf 92}, 090402 (2004).

\bibitem{brack} M. Brack and R.K. Bhaduri, {\it Semiclassical Physics}, Addison-Wesley, Reading, MA (1997).

\bibitem{george} G.F. Bertsch and S.-Y. Chang, Phys. Rev. A {\bf 76}, 021603(R) (2007)

\bibitem{urban} M.\ Grasso and M.\ Urban, Phys. Rev. A {\bf 68},033610 (2003).

\bibitem{grimm} M. Bartenstein \etal Phys. Rev. Lett. {\bf 94}, 103201 (2005).

\bibitem{ts} T. Sch\"affer, cond-mat/0701251.

\bibitem{sound} J. Kinast, A. Turlapov, and J.E. Thomas, Phys. Rev. Lett. {\bf 94} 170404 (2005).



\end{thebibliography}
\end{document}